\documentclass[prd,a4paper,showpacs,byrevtex,twocolumn]{revtex4} 
\usepackage{graphicx}
\usepackage{dcolumn}
\usepackage{amsmath}
\usepackage{array}
\usepackage{bm}
\usepackage{textcomp}

\begin{document}

\title{Hybrid meson masses and the correlated Gaussian basis}

\author{Vincent \surname{Mathieu}}
\email[E-mail: ]{vincent.mathieu@umh.ac.be}
\affiliation{Groupe de Physique Nucl\'{e}aire
Th\'{e}orique, Universit\'{e} de Mons, Acad\'{e}mie universitaire Wallonie-Bruxelles,
Place du Parc 20, BE-7000 Mons, Belgium}

\date{\today}

\begin{abstract}
We revisited a model for charmonium hybrid meson with a magnetic gluon [Yu.~S.~Kalashnikova
and A.~V.~Nefediev, Phys.\ Rev.\ D {\bf 77}, 054025 (2008)] and improved the numerical
calculations. These improvements support the hybrid meson interpretation of $X(4260)$. Within
the same model, we computed the hybrid meson mass with an electric gluon which is resolved to
be lighter. Relativistic effects and coupling channels decreased also the mass.
\end{abstract}


\maketitle
\section{Introduction}
Quantum Chromodynamics (QCD) is widely believed to be the theory of strong interactions. In
this non Abelian theory, the gauge bosons, the gluons, carry a color charge. We are therefore
led to observable color singlet configurations made of quarks but also gluons. Glueballs,
bound states of only gluons, and hybrid mesons are consequences and predictions of QCD.
Besides glueballs, hybrid mesons deserve much interest. The gluonic excitation leads to
low-lying states with quantum numbers not allowed for usual mesons (Note that the low-lying
gluon states are non exotic but glueballs with exotic quantum number also exits and are called
oddballs, for a review see Ref.~\cite{Mathieu:2008me}). Their observation would be another great
confirmation of QCD. However, the detection of exotic hadrons remains a challenging task (for
a review see Ref.~\cite{Klempt:2007cp}).

The properties of hybrid mesons were investigated in various approaches, see for examples Ref.
\cite{Isgur:1984bm,General:2006ed}, and in lattice QCD~\cite{Juge:1997nc}. The former have two
common interpretations. In the first one, the flux tube linking the quark to the antiquark is
in an excited state, allowing quantum numbers that cannot be reached by the usual
quark-antiquark picture. In the other scenario, the excitation is modelled by a constituent
gluon, leading to a three-body system. We know that both picture are closely linked: The
constituent gluon creates an equivalent potential compatible with the energy of an excited
string \cite{Buisseret:2006sz,Szczepaniak:2005xi}.

All these approaches rely on quasiparticle interpretation. For heavy particles, the spectrum
and decay properties are extracted from a Hamiltonian formalism. One has then to resort to a
numerical procedure to find eigenvalues and eigenfunctions. For two-body systems, like heavy
charmonia, different methods exist. The most efficient one is the so called Lagrange mesh
method. This technique allows one to compute energies and wave functions straightforwardly
since the method is not variational~\cite{baye86}. Matrix elements for semi-relativistic
kinetic energy are easily computed. However, it does not admit a three-body generalization in
the semi-relativistic case. For three-body systems, like hybrid mesons, one generally computes
the mass within a finite dimensional basis. We expand the unknown wave function on a set of
trial functions forming a basis of the Hilbert space in the limit where the dimension of the basis goes to infinity.

In practice, we only deal with a finite number of basis functions. Fortunately, we know the
approximation of the mass to be always an upper of the true eigenvalue, allowing to minimize
the mass with respect to parameters \cite{suzu98}. The question is then: How many basis
functions should we consider to have an acceptable accuracy ? In this work, we answer to this
question for the three-body description of hybrid mesons.

Recently, Kalashnikova and Nefediev investigated the spectrum of the charmonium hybrid $c\bar
cg$ within the constituent gluon model \cite{Kalashnikova:2008qr}. They derived a Hamiltonian
thanks to the field correlator method and introduced einbein (or auxiliary) fields to deal
with relativistic kinematics. The resulting Hamiltonian was diagonalized  with one trial
function that was taken as a Gaussian depending of a hyperradius. Finally, the authors
computed correction coming from the string, self-energy and spin-dependent operators. The
authors studied the case of the lowest $c\bar cg$ hybrid meson in with the gluon quantum
numbers are $\ell_g=1$ and $j=1$ (magnetic gluon). In this work, we investigate how accurate
are those approximations for the wave function (truncation of the basis) and for the Hamiltonian (introduction of auxiliary fields) and greatly improve the accuracy on the mass thanks to correlated Gaussian functions.

In Sec.~\ref{sec:def}, we present the model of Ref.\cite{Kalashnikova:2008qr} and recall its
main properties. The authors used a particular ansatz for the wave wave function which is
described in Sec.~\ref{sec:hyperS}. We discuss some improvements for the wave function in
Sec.~\ref{sec:Gauss}. In this section, which consist in the main part of this work, we perform numerical calculations with various approximations for the model and make some comparisons.
Having identified a good approximation for the wave function, we test in Sec.~\ref{sec:param}
the stability of the error under parameter evolution. Finally, we draw our conclusion in
Sec.~\ref{sec:conclu}

\section{Hybrid mesons as $q\bar qg$ systems} \label{sec:def}
The starting point for the hybrid model of \cite{Kalashnikova:2008qr} is a Hamiltonian derived
from QCD thanks to the field correlator method. For a hybrid meson seen as a $q\bar q g$
system, this Hamiltonian reads
\begin{equation}\label{eq:mainH}
    H=H_0 +  V_{\text{C}},
\end{equation}
where
\begin{equation}
\begin{split}
  H_0= & \sqrt{\bm p^2_q+m^2_q}+\sqrt{\bm p^2_{\bar q}+m^2_{\bar q}} +\sqrt{\bm p^2_g} \\
    &+\sigma|\bm x_g-\bm x_q|+\sigma|\bm x_g-\bm x_{\bar q} |,
\end{split}
\end{equation}
\begin{equation}\label{eq:coul}
V_{\text{C}}=-\frac{3\alpha_s}{2|\bm x_g-\bm x_q|}-\frac{3\alpha_s}{2|\bm x_g- \bm x_{\bar
q}|} + \frac{\alpha_s}{6|\bm x_q- \bm x_{\bar q}|}.
\end{equation}
The first terms of $H_0$ represent the kinetic energy of two quarks of the same mass and of
the massless constituent gluon. The long-range confining force consists in two strings, two
fundamental flux tubes of energy density $\sigma$, linking the gluon to the quarks. Such a
confining potential is in agreement with other phenomenological
approaches~\cite{horn,Mathieu:2005wc} as well as with recent lattice QCD
computations~\cite{bicu}. For other three-body systems like baryons, the confinement is generally assumed to be a Y-shape. In the case of the hybrid, however, under Casimir scaling hypothesis for the string tension, the V-configuration is resolved to be energetically favorable \cite{Mathieu:2005wc}. The Torricelli point of the Y-shape merges with the gluon position. Imposing a Y-shape would obviously raise the whole spectrum. The Casimir scaling hypothesis is supported by recent lattice studies, see for instance ref.~\cite{Bali:2000un}.

The confining potential is augmented by the short-range Coulomb potential $V_{\text{C}}$, where $\alpha_s$ is the strong coupling constant, arising from one gluon exchange effects between the constituent particles. In particular, the Coulomb potential between the quark and the antiquark is repulsive since the $q\bar q$ pair is in a color octet. We refer the reader to Ref.~\cite{oge} for a detailed discussion about the short-range interaction potentials in QCD. Spin-dependent short-range interactions could be added in perturbation as it is done in
Ref.~\cite{Kalashnikova:2008qr}. However, it is enough for our purpose to consider only the
dominant order Hamiltonian~(\ref{eq:mainH}).

To be complete, we recall that the parity and charge conjugation of a $q\bar q g$ system are
given respectively by
\begin{align}
    P&=(-1)^{\ell_{q\bar q}+\ell_g}, & C &= (-1)^{1+\ell_{q\bar q}+S_{q\bar q}},
\end{align}
where $\ell_{q\bar q}$ and $\ell_g$ are the orbital angular momenta of the $q\bar q$ pair and
of the gluon, and where $S_{q\bar q}$ is the intrinsic spin of the $q\bar q$ pair. Compared to
usual mesons, extra phases give rise to exotic quantum numbers such as $1^{-+}$ for example.
Two cases have to be distinguished following the total spin $j$ of the gluon: The magnetic
gluon, for which $\ell_g=j$, and the electric gluon for which
$\ell_g=j\pm1$~\cite{Kalashnikova:2008qr}. Although it may gain a dynamical mass induced by
confinig forces, the gluon is a massless particle that remains transverse, with only two
polarisations~\cite{Mathieu:2008bf,Mathieu:2008me}. Therefore, the minimal value for $j$ is 1.
Exotic quantum numbers require at least one $P$-wave. In the following, we shall refer to
magnetic (electric), the $q\bar q g$ states with $\ell_g=1$ ($\ell_g=0$).

Solving the eigenequation associated to a three-body semirelativistic Hamiltonian such
as~(\ref{eq:mainH}) is a difficult numerical problem. The intrinsic complexity of three-body
systems obviously comes into play, but another problem is the determination of the matrix
elements for kinetic operators of the form $\sqrt{\bm p^2+m^2}$, which are not commonly found
in quantum mechanics. This last difficulty can be avoided by introducing auxiliary (or
einbein) fields to get rid of the square roots appearing in Hamiltonian~(\ref{eq:mainH}). One
obtains
\begin{equation}\label{eq:mainH_AF}
\begin{split}
  H(\mu_q,\mu_{\bar q},\mu_g)&= \frac{\mu_q+\mu_{\bar q}+\mu_g}{2}+ \frac{\bm p^2_q+m^2_q}{2\mu_q}+  \frac{\bm p^2_{\bar
q}+m^2_{\bar q}}{2\mu_{\bar q}}  \\
    & + \frac{\bm p^2_g}{2\mu_g} + \sigma|\bm x_g-\bm
x_q|+\sigma|\bm x_g-\bm x_{\bar q} | + V_{\text{C}},
\end{split}
\end{equation}
where $\mu_q$, $\mu_{\bar q}$, and $\mu_g$ are the so-called auxiliary fields. Being formally
defined as operators, they can be eliminated through the following equations
\begin{align}
    \left.\delta_{\mu_q} H(\mu_q,\mu_{\bar q},\mu_g)\right|_{\mu_q=\hat\mu_q}=
    0\Rightarrow \hat\mu_q&=\sqrt{\bm p^2_q+m^2_q},\\
    \left.\delta_{\mu_{\bar q}} H(\mu_q,\mu_{\bar q},\mu_g)\right|_{\mu_{\bar q}=
    \hat\mu_{\bar q}}=0\Rightarrow \hat\mu_{\bar q}&=\sqrt{\bm p^2_{\bar q}+m^2_{\bar q}},\\
\left.\delta_{\mu_g} H(\mu,\mu_{\bar q},\mu_g)\right|_{\mu_g=\hat\mu_g}=0\Rightarrow
\hat\mu_g&=\sqrt{\bm p^2_g}.
\end{align}
It is then readily checked that $H(\hat\mu_q,\hat\mu_{\bar q},\hat\mu_g)=H_0$; both
Hamiltonians are equivalent up to the elimination of the auxiliary fields as operators.
However, the calculations are considerably simplified if one considers them are c-numbers
variational parameters. The eigenvalues of the spinless Hamiltonian \eqref{eq:mainH_AF} are
more easily found since only nonrelativistic kinetic operators are present. These eigenvalues,
denoted as $E_0(\mu_q,\mu_{\bar q},\mu_g)$, are finally minimized with respect to the
einbeins. The optimal values of the auxiliary fields seen as variational parameters are
logically close to the average values of the corresponding operators~\cite{lagmesh,Af1}. For
example, the optimal value of $\mu_g$, denoted as $\mu_{g0}$, is such that
$\mu_{g0}\approx\left\langle \hat\mu_g\right\rangle$. It can be interpreted as a dynamical
gluon mass. The same arguments hold for the other auxiliary fields. An important point, that
has been shown in Ref.~\cite{hybaf}, is that the eigenvalues of
Hamiltonian~(\ref{eq:mainH_AF}) are upper bounds of the eigenvalues of
Hamiltonian~(\ref{eq:mainH}). The more auxiliary fields are introduced, the less this bound is
accurate. In particular, the accuracy of the auxiliary field method decreases when light or
massless particles are present in the system under study.

In this work we focus on the case where the quark and the antiquark have the same mass, {\it
i.e.} $m_q=m_{\bar q}=m$. Then, by symmetry, $\mu_q=\mu_{\bar q}=\mu$ and, using the Jacobi
coordinates
\begin{equation}
    \bm \lambda=\bm x_q-\bm x_{\bar q},\quad \bm \rho=\bm x_g-\frac{\bm x_q+\bm x_{\bar q}}{2},
\end{equation}
Hamiltonian~(\ref{eq:mainH_AF}) becomes
\begin{equation}\label{AF3}
\begin{split}
  H(\mu,\mu_g)= &  \mu+\frac{\mu_g}{2}+\frac{m^2}{\mu}+\frac{\bm p^2_\lambda}{\mu}+ \frac{\bm p^2_\rho}{2\phi}\\
    & +\sigma\left| \frac{\bm\lambda}{2}-\bm \rho\right|+\sigma\left|\frac{\bm \lambda}{2}+\bm
\rho\right|+ V_{\text{C}},
\end{split}
\end{equation}
where $\bm p_\lambda$ and $\bm p_\rho$ are the momenta associated to $\bm\lambda$ and
$\bm\rho$ respectively, and where
\begin{equation}
    \phi=\frac{2\mu\mu_g}{2\mu+\mu_g}.
\end{equation}
The center of mass, defined as
\begin{equation}
    \bm R=\frac{\mu\bm x_q+\mu\bm x_{\bar q}+\mu_g\bm x_g}{2\mu+\mu_g},
\end{equation}
is decoupled and its conjugate momentum is set equal to zero since we work in the rest frame
of the system. 

\section{Hyperspherical formalism}\label{sec:hyperS}

Eigenvalues of Hamiltonian $H(\mu,\mu_g)$ with auxiliary fields~(\ref{AF3}) in the case of
$c\bar c g$ systems have been numerically computed in Ref.~\cite{Kalashnikova:2008qr} by using
the hyperspherical formalism. We recall in this section the procedure that has been used in
this last reference. The authors of \cite{Kalashnikova:2008qr} state that since the $q\bar q$
pair is heavy, the assumption $\ell_{q\bar q}=0$ can be made. Moreover, the particular case of
a magnetic gluon with $\ell_g=1$ is considered in this reference. For this particular choice,
the spin of the quark pair $S_{q\bar q}$ is a good quantum number and leads to the following
states: $1^{--}$ for $S_{q\bar q}=0$ and $(0,1,2)^{-+}$ for $S_{q\bar q}=1$. All these states
are degenerate since $H(\mu,\mu_g)$ does not have any spin-dependent term. This leads to wave
functions of the form
\begin{equation}\label{}
\left|\chi(\bm X) \right\rangle\otimes\left[ \left|\ell_{q\bar
q}=0,\ell_g=1\right\rangle^1\otimes \left|S_{q\bar q}\right\rangle\right]^J,
\end{equation}
with the hyperradius
\begin{equation}\label{hyperradius}
    \bm X^2=\frac{\mu}{2}\bm\lambda^2 +
\frac{2\mu\mu_g}{2\mu+\mu_g}\bm\rho^2.
\end{equation}
$\left|\ell_{q\bar q},\ell_g\right\rangle^\ell$ is a shorthand notation for the coupling
$\left[{\cal Y}_{\ell_{q\bar q}}(\bm\lambda)\otimes{\cal Y}_{\ell_g}(\bm\rho)\right]^\ell$,
with the solid spherical harmonics ${\cal Y}_{\ell_gm}(\bm \rho)=\rho^{\ell_g}
Y_{\ell_gm}(\hat{\bm\rho})$. The following states can consequently be described within this
approach
\begin{eqnarray}
    \left|1^{--}\right\rangle&=&\left|\chi(\bm X) \right\rangle\otimes\left[\left|0,1\right\rangle^1\otimes
    \left|0\right\rangle\right]^1\\
\left|J^{-+}\right\rangle&=&\left|\chi(\bm X)
\right\rangle\otimes\left[\left|0,1\right\rangle^1\otimes \left|1\right\rangle\right]^{J},
\end{eqnarray}
with $J=\{0,1,2\}$. The trial wave function is chosen to be a Gaussian depending on one
variational parameter $\beta$, {\it i.e.}
\begin{equation}\label{eq:trialfct_hyperS}
\left\langle \bm\rho,\bm\lambda \right|\left[\left.\chi(\bm
X)\right\rangle\otimes\left|0,1\right\rangle^1\right]
 = \exp\left(-\frac{1}{2}\beta^2\bm X^2\right){\cal Y}_{1m}(\bm \rho).
\end{equation}
This solid harmonic, ${\cal Y}_{1m}(\bm \rho)=\rho Y_{1m}(\hat{\bm\rho})$, determines the
angular momentum in order to treat exotic mesons with a $P$-wave gluon and a $S$-wave $q\bar
q$ pair. The spin function for the quark pair is irrelevant since our Hamiltonian is
spin-independent. We will therefore get the same mass for the the four states
$0^{-+},1^{--},1^{-+},2^{-+}$ under consideration.

A convenient method to fit the parameters is to reproduce the charmonium spectrum within the
same flux tube model, that is with the Hamiltonian $H_{c\bar  c}=2\sqrt{\bm p^2+m^2_c}+\sigma
r-4\alpha_s/3r$. The parameters obtained in this way are compatible with typical values used
in potential models~\cite{Kalashnikova:2008qr}:
\begin{align}\label{eq:param}
    \sigma&=0.16\text{~GeV$^2$,} & \alpha_s&=0.55,& m_c&=1.48 \text{~GeV}.
\end{align}
We, as the authors of ref.~\cite{Kalashnikova:2008qr}, then implicitly assumed the same value for the parameters for both systems (charmonium and hybrid charmonium). It is also natural to keep the same parameters as in ref.~\cite{Kalashnikova:2008qr} since the present paper focuses on the improvement of the numerical resolution. In Sec.~\ref{sec:param} we vary the parameters and investigate this influence of the one Gaussian approximation.

A numerical resolution of $H(\mu,\mu_g)$ with the ansatz \eqref{eq:trialfct_hyperS} leads
to~\cite{Kalashnikova:2008qr}
\begin{align}\label{eq:resultbad}
    M_0&=4.573\text{~GeV,} & \mu_0&=1.598\text{~GeV,} & \mu_{g0}&=1.085\text{~GeV},
\end{align}
for the ground state. This mass corresponds in a first approximation to $c\bar c$ hybrid
mesons with quantum numbers $1^{--}$ and $(0,1,2)^{-+}$.

Spin-dependent corrections have also been computed to be equal to~\cite{Kalashnikova:2008qr}
\begin{subequations}\label{splittings}
\begin{eqnarray}
  \Delta M(0^{-+})&=&-321 \text{ MeV,} \\
  \Delta M(1^{-+})&=&-253 \text{ MeV,} \\
  \Delta M(1^{--})&=&-176 \text{ MeV,} \\
  \Delta M(2^{-+})&=&-116 \text{ MeV.}
\end{eqnarray}
\end{subequations}
The splittings induced by the spin-dependent operators in perturbation theory are clearly not
negligible. However, we know from the Rayleigh-Ritz method \cite{suzu98} that the mass
$M_0=4.573$ GeV can only be an upper bound of the true eigenvalue of $H(\mu,\mu_g)$. Indeed,
when truncating the basis with a finite number of basis function, one restrict the operator to the subspace spanned by the trial functions. It is then natural to investigate how far is
$M_0$ from the true eigenvalue and compare the difference with respect to the additional
corrections \eqref{splittings}.

\section{Correlated Gaussian basis}\label{sec:Gauss}
The real eigenvalue $M$ of a Schr\"{o}dinger equation $H|\Phi\rangle=M|\Phi\rangle$ is
generally found by expanding the wave function $\Phi(\bm\lambda,\bm\rho) =\langle\bm
\rho,\bm\lambda|\Phi\rangle$ in a basis
\begin{equation}\label{}
\Phi(\bm\lambda,\bm\rho) = \sum_i^N\alpha_i\varphi^i(\bm\lambda,\bm\rho).
\end{equation}
The real eigenvalue is reached in the limit $N\to\infty$. Since for fixed $N$, $M^{(N)}$ is an
upper bound, we can then minimize the mass with respect to the parameters on which depend
$\varphi^i$. A appropriate ansatz for $\varphi^i$ will lead to an accurate mass for low values
of $N$.

For the trial functions, we use a generalisation of the hyperspherical function
\eqref{eq:trialfct_hyperS}
\begin{equation}\label{eq:def:Gaussian}
\varphi^i(\bm\lambda,\bm\rho) =
\exp\left(-a_i\bm\lambda^2-b_i\bm\rho^2-2c_i\bm\lambda\cdot\bm\rho\right) {\cal
Y}_{LM}(d_i\bm\lambda+e_i\bm\rho).
\end{equation}
The wave function used in \cite{Kalashnikova:2008qr} is recovered for the particular values
$N=1$, $c_1=d_1=0$ and the condition \eqref{hyperradius} on the variational parameters $a_1$
et $b_1$. The correlated Gaussian \eqref{eq:def:Gaussian} shared many advantages
\cite{suzu98}. They admit a generalisation for an arbitrary number of interacting particles
thanks to matrix notation; matrix elements for usual power laws potential are expressed in
close forms; the Fourier transform of a Gaussian is also a Gaussian. In particular, we have
shown that matrix elements for semirelativistic kinematics $\sqrt{\bm p^2+m^2}$ can be easily
derived \cite{Fourier,LS_tenseur}.

This form is very convenient for hybrid mesons since we deal easily with angular excitations
for the gluon but also for the quarks. Imposing some restriction on the parameters we get two
ansatz for the lowest $P$-wave hybrid mesons
    \begin{align}\label{eq:gauss_gluon}
    \varphi_A(\bm\lambda,\bm\rho) &=\exp\left(-a\bm\lambda^2-b\bm\rho^2\right){\cal Y}_{1
    m}(\bm\rho),\\
    \label{eq:gauss_QQ}
        \varphi_B(\bm\lambda,\bm\rho) &= \exp\left(-a\bm\lambda^2-b\bm\rho^2\right)
        {\cal Y}_{1m}(\bm\lambda).
    \end{align}
The function  $\varphi_A(\bm\lambda,\bm\rho)$ has the angular part $|\ell_{q \bar
q}=0,\ell_g=1\rangle^1$ and correspond to a hybrid meson with a magnetic gluon,  and the
function $\varphi_B(\bm\lambda,\bm\rho) $ has the angular behavior $|\ell_{q \bar
q}=1,\ell_g=0\rangle^1$ and corresponds to a hybrid meson with an electric gluon.

\subsection{$P$-wave hybrids with magnetic gluon}

\begin{figure}[htb]\begin{center}
\resizebox{0.5\textwidth}{!}{
  \includegraphics*{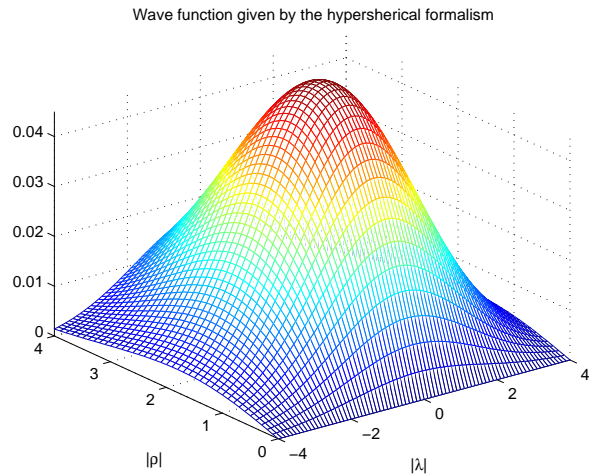}
  }\caption{\label{fig:A:Hyper}Wave function \eqref{eq:trialfct_hyperS} for a magnetic gluon hybrid meson.}
\end{center}
\end{figure}

\begin{figure}[htb]\begin{center}
\resizebox{0.5\textwidth}{!}{
  \includegraphics*{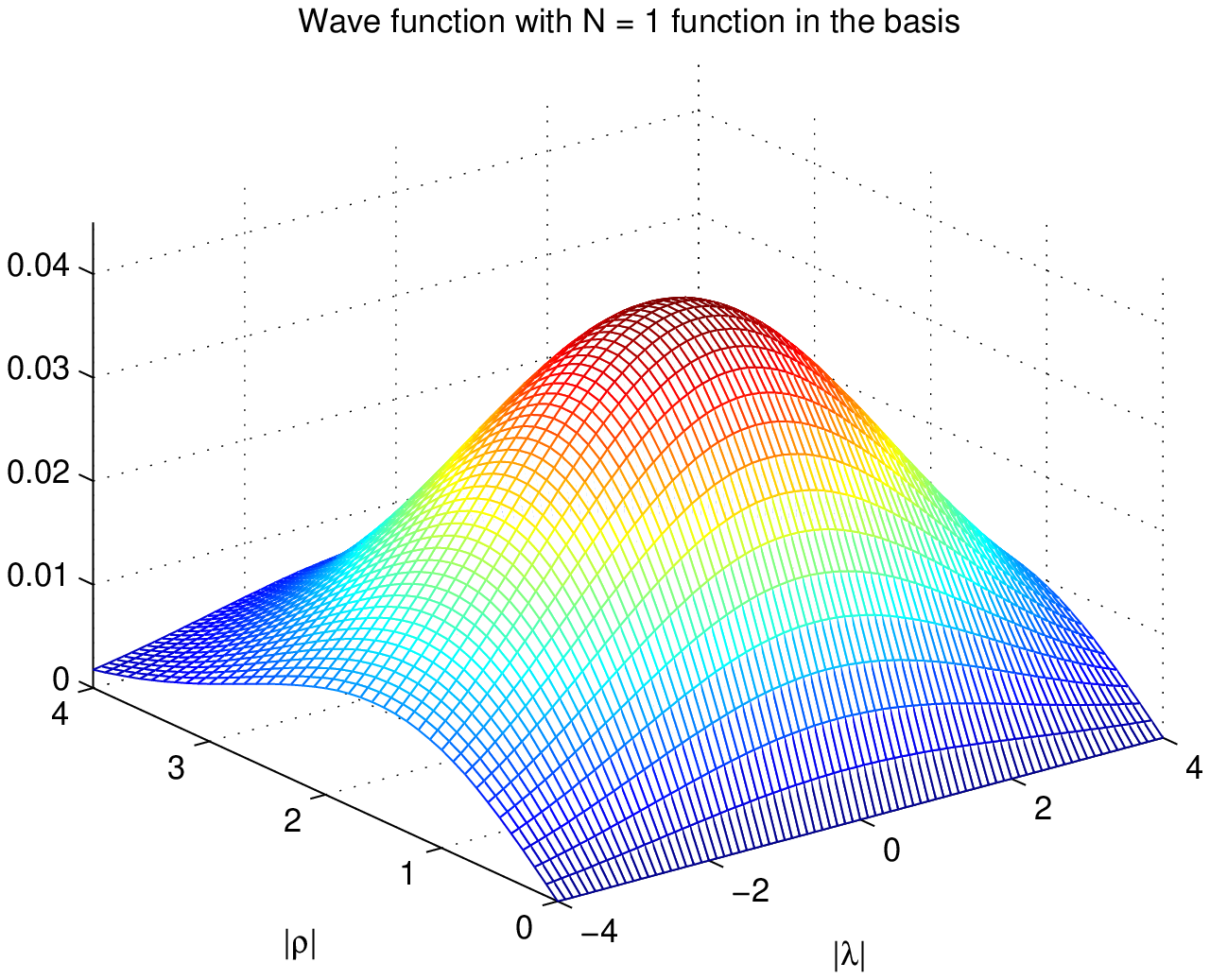}
  }\caption{\label{fig:A:one}Wave function \eqref{eq:gauss_gluon} for a magnetic gluon hybrid meson.}
\end{center}
\end{figure}

Let us now see how changes the mass when relaxing the condition \eqref{hyperradius} on $a_1$
and $b_1$. In order to have a relevant comparison point, we consider $H(\mu,\mu_g)$ with the
parameters given in \eqref{eq:param} and \eqref{eq:resultbad}. If we use the hyperspherical
formalism with only one trial function (and hence only one variational parameter) in our
basis, we find a mass of $M_0 = 4.573$ GeV (obviously the same value as in
Ref.~\cite{Kalashnikova:2008qr}). The optimisation procedure leads to $\beta=0.62$ GeV$^{1/2}$
which correspond to $a^{H}=0.154$ GeV and $b^{H}=0.156$ GeV. The relation \eqref{hyperradius}
with the parameters \eqref{eq:param} implies a nearly symmetric Gaussian shape, {\it i.e}
\begin{equation}\label{eq:ratio_Hyper}
    \frac{b^H}{a^H} = \frac{2\mu+\mu_g}{4\mu_g}=1.0138.
\end{equation}
Here we perform a numerical optimisation on the two parameters of our single wave function
$\varphi_A(\bm\lambda,\bm\rho)$. We expect to find a mass slightly lower than in
\cite{Kalashnikova:2008qr}. Indeed, the resulting mass for the four states
$0^{-+},1^{--},1^{-+},2^{-+}$  in this approximation reads
\begin{equation}\label{}
    M^{(1)}_{A} = 4.462 \text{ GeV}.
\end{equation}
We gain 109 MeV by simply choosing a more general ansatz (with two variational parameters
instead of one parameter) for the trial wave function. The shift is of order of the
spin-splittings \eqref{splittings}. Let us look at the wave in order to understand this
difference. The optimisation procedure leads to $a_1=0.083$ GeV and $b_1=0.204$ GeV. Their
ratio strongly differs from \eqref{eq:ratio_Hyper}. As an illustration, we plot the spatial
part of the wave functions in Figs \ref{fig:A:Hyper} and \ref{fig:A:one} (remind the factor
$\rho$ in the solid spherical harmonic).


\begin{table} [h]
\begin{ruledtabular}\caption{Wave function coefficients for magnetic gluon \eqref{eq:gauss_gluon}.}
  \begin{tabular}{cccc}
    $i$ & $\alpha_i$ $10^{-3}$& $a_i$ & $b_i$ \\
    \hline
    1 &26.183 &0.082811 &0.203578  \\
    2 &116.410 &0.207870 &0.697459  \\
    3 &$-6.902$ &0.435376 &0.232531  \\
    4 &$-3.242$ &2.429518 &0.720520  \\
    5 &$-4.284$ &0.050134 &0.088911  \\
    6 &$-11.033$ &0.079373 &1.083885  \\
    7 &46.785 &0.153637 &0.383596  \\
    8 &8.5083 &0.053970 &0.097767  \\
    9 &$-0.505$ &0.165116 &0.078640  \\
    10 &$-98.834$ &0.208914 &0.571826  \\
  \end{tabular}
  \label{tab:wfmgt}
\end{ruledtabular}
\end{table}

\begin{figure}[htb]\begin{center}
\resizebox{0.5\textwidth}{!}{
  \includegraphics*{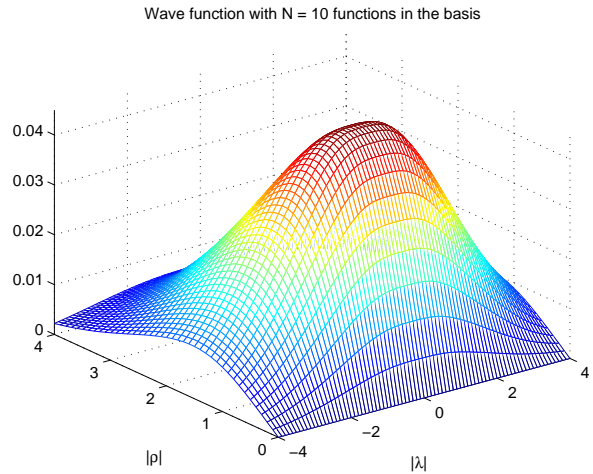}
  }
\caption{\label{fig:A:ten}Wave function \eqref{eq:def:Gaussian} with 10 Gaussian functions
\eqref{eq:gauss_gluon} for a magnetic gluon hybrid meson.}
\end{center}
\end{figure}

We can go even further by adding more functions in the basis. With a sufficient number of
functions we should converge to the real eigenvalue of our operator $H(\mu,\mu_g)$.
Convergence is obtained for a small number of Gaussians. With $N=10$ functions, the numerical
procedure is accurate up to 1 MeV which is clearly enough. The wave functions is then a sum of 10 Gaussian trial functions. The coefficients and parameters of each functions can be read in Table~\ref{tab:wfmgt}. The resulting mass
\begin{equation}\label{}
M^{(10)}_A=4.445 \text{ GeV}
\end{equation}
is somewhat lower. The gain when increasing the basis is not as strong as the first gain
obtained with one trial Gaussian \eqref{eq:gauss_gluon} with two variational parameters. We
can understand this effect as follows: The Gaussian wave function \eqref{eq:gauss_gluon} has a good overlap with the exact wave function of the three-body system. However, when we restrict
the trial function to be a Gaussian of the only hyperradius \eqref{eq:trialfct_hyperS}, the
overlap with the wave function is worst since the hyperradius imposes a relation relation
between the two coefficients of Jacobi variables which is not realised in the optimal case.
The wave functions with $N=10$ functions is shown in Fig.~\ref{fig:A:ten}. The relative error
between the first trial function \eqref{eq:trialfct_hyperS} and the eigenfunction is $33\%$
and the relative error ({\it i.e.} the ratio between the integral of the difference squared
and the integral of the function with $N=10$ squared) with the single Gaussian approximation
is $18\%$. Nevertheless, the mass found with one hyperspheric Gaussian is less than $3\%$
above the real value.

In their paper, the authors of \cite{Kalashnikova:2008qr} concluded favorably in a hybrid
meson interpretation of the vector meson candidate $X(4260)$. This new state was observed by the BaBar collaboration~\cite{Aubert:2005rm}. But they mentioned also that
their mass $M_0+\Delta M(1^{--}) = 4.397$ GeV was higher than the candidate's mass. It is
worth mentioning that the exact value of the Hamiltonian $H(\mu,\mu_g)$ used in
\cite{Kalashnikova:2008qr} is 130 MeV below the approximative value computed with the trial
function \eqref{eq:trialfct_hyperS}. Assuming that spin-dependent corrections do not change
much with the change of the wave function, we get a mass compatible with the $X(4260)$.

\subsection{$P$-wave hybrids with electric gluon}
In the previous section, we focused on the magnetic gluon hybrids. We followed the approach of \cite{Kalashnikova:2008qr} where the authors assumed a $S$-wave quark-antiquark pair. Since,
the quark-antiquark pair is a repulsive octet state, one can then wonder if it would not be
energetically favourable to consider a $P$-wave for the $q\bar q$ instead of exciting the
gluon. The trial function respecting this assumption is $\varphi_B(\bm\rho,\bm\lambda)$ in
\eqref{eq:gauss_QQ}. This ansatz ($\ell_{q\bar q}=1$), not investigated in
Ref.~\cite{Kalashnikova:2008qr}, corresponds to an electric gluon hybrid with quantum numbers
$J^{--},J\in[0,3]$ or $J^{-+},J\in[0,2]$.

First, we used only one single trial function depending on the hyperradius. We considered the
same Hamiltonian $H$ with the same value for $\sigma,\alpha_s,m_c,\mu$ and $\mu_g$. We found a
mass $M_0'=4.225$ GeV which is already lower than the mass of a magnetic gluon hybrid. This
time, the variational parameter has the value $\beta=0.633$ GeV$^{1/2}.$

\begin{figure}[htb]\begin{center}
\resizebox{0.5\textwidth}{!}{
  \includegraphics*{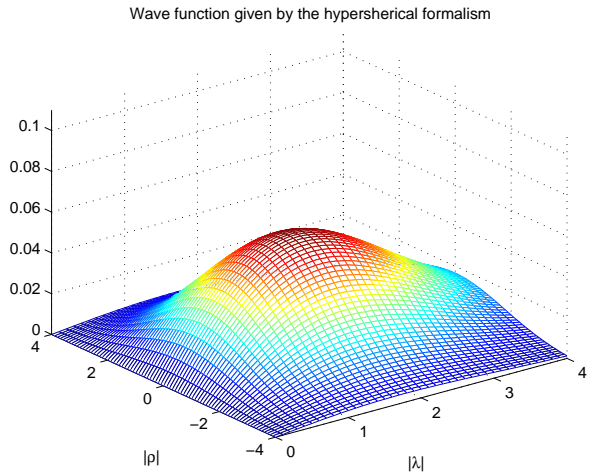}
  }\caption{\label{fig:B:Hyper}Wave function \eqref{eq:trialfct_hyperS} for an electric gluon hybrid meson.}
\end{center}
\end{figure}

\begin{figure}[htb]\begin{center}
\resizebox{0.5\textwidth}{!}{
  \includegraphics*{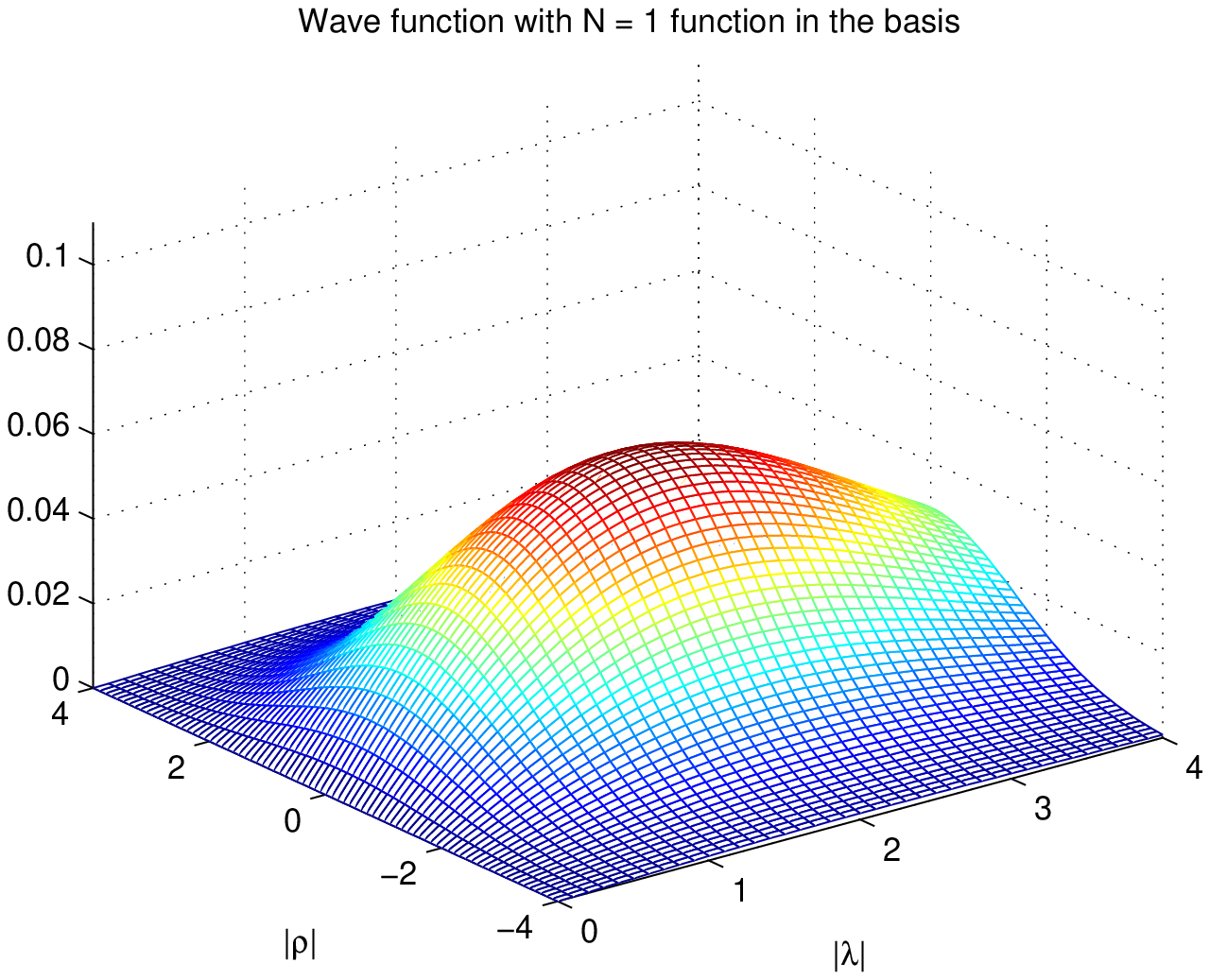}
  }\caption{\label{fig:B:one}Wave function \eqref{eq:gauss_gluon} for an electric gluon hybrid meson.}
\end{center}
\end{figure}

As a next step, we diagonalized $H(\mu,\mu_g)$ with one Gaussian trial function with two
parameters. We found a mass $M_B^{(1)} = 4.137$ GeV with only one Gaussian function
$\varphi_B(\bm\rho,\bm\lambda)$. The optimal values for the parameters read: $a_1=0.119$ GeV
and $b_1=0.232$ GeV. As in the case of a magnetic gluon, we found a mass roughly 100 MeV lower
than with one function with the hyperspherical formalism. We plot the spatial part of the wave
functions in Figs \ref{fig:B:Hyper} and \ref{fig:B:one} (remind the factor $\lambda$ in the
solid spherical harmonic).


\begin{table} [h]
\begin{ruledtabular}\caption{Wave function coefficients for electric gluon}
  \begin{tabular}{cccc}
    $i$ & $\alpha_i$ $10^{-3}$& $a_i$ & $b_i$ \\
    \hline
    1 &38.013 &0.119838 &0.232252  \\
    2 &0.948 &0.051007 &0.087008  \\
    3 &$-0.977$ &0.039315 &0.393460  \\
    4 &5.333 &0.319606 &0.714934  \\
    5 &$-18.77$ &0.091608 &1.320736  \\
    6 &$-5.593$ &0.398742 &0.209574  \\
    7 &9.742 &0.070955 &1.500372  \\
    8 &55.451 &0.257114 &0.710774  \\
    9 &2.330 &0.066225 &0.219531  \\
    10 &34.224 &1.050299 &2.419412  \\
  \end{tabular}
  \label{tab:wfelec}
\end{ruledtabular}
\end{table}

\begin{figure}[htb]\begin{center}
\resizebox{0.5\textwidth}{!}{
  \includegraphics*{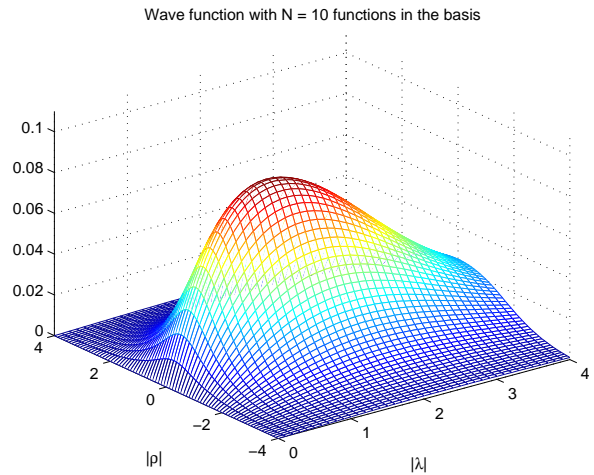}
  }
\caption{\label{fig:B:ten}Wave function \eqref{eq:def:Gaussian} with 10 Gaussian functions
\eqref{eq:gauss_gluon} for an electric gluon hybrid meson.}
\end{center}
\end{figure}

The next step is obviously to check how accurate is the one Gaussian approximation by looking
for the exact eigenvalue of $H(\mu,\mu_g)$ (by exact, we mean a value stable at 1 MeV). When
increasing $N$, the dimension of our basis, we reach a stable mass from $N=10$. We check up to $N=30$ the stability of our eigenvalue. The mass of electric gluon hybrid reads
\begin{equation}\label{}
    M_{B}^{(10)} = 4.115 \text{ GeV}.
\end{equation}
The coefficients and parameters of each functions can be read in Table~\ref{tab:wfelec}.
Once again, the difference between the simple Gaussian and the true is five time lower than
the difference between the two first approximations. The wave functions for $N=10$ is shown in
Fig.~\ref{fig:B:ten}. The relative error between the first trial function
\eqref{eq:trialfct_hyperS} and the eigenfunction is $29\%$ and the relative error with the
single Gaussian approximation is $26\%$.

We then conclude that the lowest exotic hybrid mesons $0^{--},1^{-+},3^{-+}$ are dominated by
the electric gluon component.

\subsection{$P$-wave hybrid meson}
We have identified two main components in the lowest $P$-wave hybrid mesons. The electric
gluon component was resolved to be lighter that the magnetic gluon component since the quarks
are in an octet state. The physical wave function should be a mixing of those two situations
since our Hamiltonian couples the two configurations. We expect then a lower mass for the
ground state since coupling two channels repels each other. The more general ansatz for a
$P$-wave hybrid meson is
\begin{equation}\label{eq:ansatzfull}
\begin{split}
\Phi(\bm\lambda,\bm\rho) =& \sum_i^N\alpha_i
\exp\left(-a_i\bm\lambda^2-b_i\bm\rho^2-2c_i\bm\lambda\cdot\bm\rho\right) \\
&\times {\cal Y}_{1M}(d_i\bm\lambda+e_i\bm\rho),
\end{split}
\end{equation}
and correspond to $0^{-+},1^{--},1^{-+},2^{-+}$. Using this ansatz, we indeed find a lowest
value for the ground state of $H(\mu,\mu_g)$:
\begin{equation}\label{}
    M = 4.068 \text{ GeV}.
\end{equation}
The resulting mass is indeed lower than in the previous cases.

\subsection{Semi-relativistic Hamiltonian}
We showed in the previous sections that the ansatz for the wave function is essential to find
the correct eigenvalues of an operator. One simple exponential of the hyperradius may lead to
a mass around 100 MeV above the true value. But, in this work, we assume that the correct
Hamiltonian for the $q\bar q g$ system is $H$ since the einbein are operators and not numbers.
As a consequence, eigenvalue of $H(\mu, \mu_g)$ are upper bounds of eigenvalues of $H$. The
auxiliary fields were introduced to get rid of the square roots in $H_0$. Treating $\mu$ and
$\mu_g$ as ordinary numbers instead of operator causes an error on the energy around
$5\%$~\cite{buis041}. Let us note also that, the more einbein there are, the less accurate is
the approximation. We can avoid this approximation by diagonalizing $H$ given by
\eqref{eq:mainH}.

The Fourier transform of the general trial function \eqref{eq:def:Gaussian} has also a
Gaussian shape. We can then easily compute matrix elements for operator in momentum space such
as the semi-relativistic kinetic operator \cite{Fourier}. With a basis of Gaussian functions,
we do not need to introduce auxiliary fields. This avoids also to determine the optimal values
of $\mu$ and $\mu_g$.

We are in position to find the accurate eigenvalues of the semi-relativistic Hamiltonian $H$
\eqref{eq:mainH}. We computed the masses of the lightest hybrid with magnetic ($\varphi_A$)
and electric ($\varphi_A$) gluons with $N=1$ and $N=10$ functions in the basis. The masses are
below the ones obtained with the auxiliary field approximation and read
\begin{align}\label{}
M_{0A}^{(1)}&=4.402 \text{ GeV,} & M_{0B}^{(1)}&=4.017 \text{ GeV}, \\
    M_{0A}^{(10)}&=4.361 \text{ GeV,} & M_{0B}^{(10)}&=3.970 \text{ GeV}.
\end{align}
The coefficients and parameters of each functions can be read in Table~\ref{tab:wfmgtsemirelat} for a magnetic gluon and in Table~\ref{tab:wfelecsemirelat} for an electric gluon. The relative difference between the real values of $H$ and $H(\mu,\mu_g)$ is less than $4\%$. With the wave functions, we computed the expectation values (for $N=20$)
\begin{align}\label{}
    \mu_0&=\langle \Phi|\sqrt{\bm p^2_q+m_c^2}|\Phi\rangle, & \mu_{g0}&=\langle \Phi|\sqrt{\bm p^2_g}|\Phi\rangle.
\end{align}
In both cases, magnetic ($A$) and electric ($B$) gluons, the expectation values of these
operators are compatible with the parameters \eqref{eq:resultbad} of
\cite{Kalashnikova:2008qr}:
\begin{align}\label{}
\mu_0^A&=1.652 \text{ GeV}, & \mu_{g0}^A&= 0.976\text{ GeV},\\
\mu_0^B&= 1.736\text{ GeV}, & \mu_{g0}^B&= 0.750\text{ GeV}.
\end{align}
As expected, $\mu_0^A<\mu_0^B$ and $\mu_{g0}^A>\mu_{g0}^B$ since in the magnetic case, the
$P$-wave is reported on the gluon.

\begin{figure}[htb]\begin{center}
\resizebox{0.5\textwidth}{!}{
  \includegraphics*{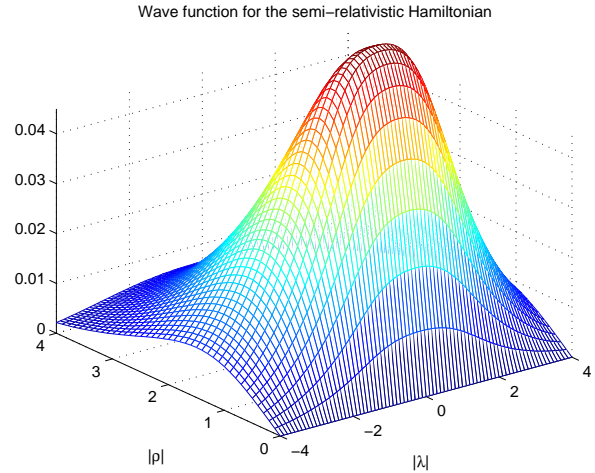}
  }
\caption{\label{fig:B:SR}Wave function for the semi-relativistic Hamiltonian
\eqref{eq:def:Gaussian} with 10 Gaussian functions \eqref{eq:gauss_gluon} for a magnetic gluon hybrid meson.}
\end{center}
\end{figure}

\begin{figure}[htb]\begin{center}
\resizebox{0.5\textwidth}{!}{
  \includegraphics*{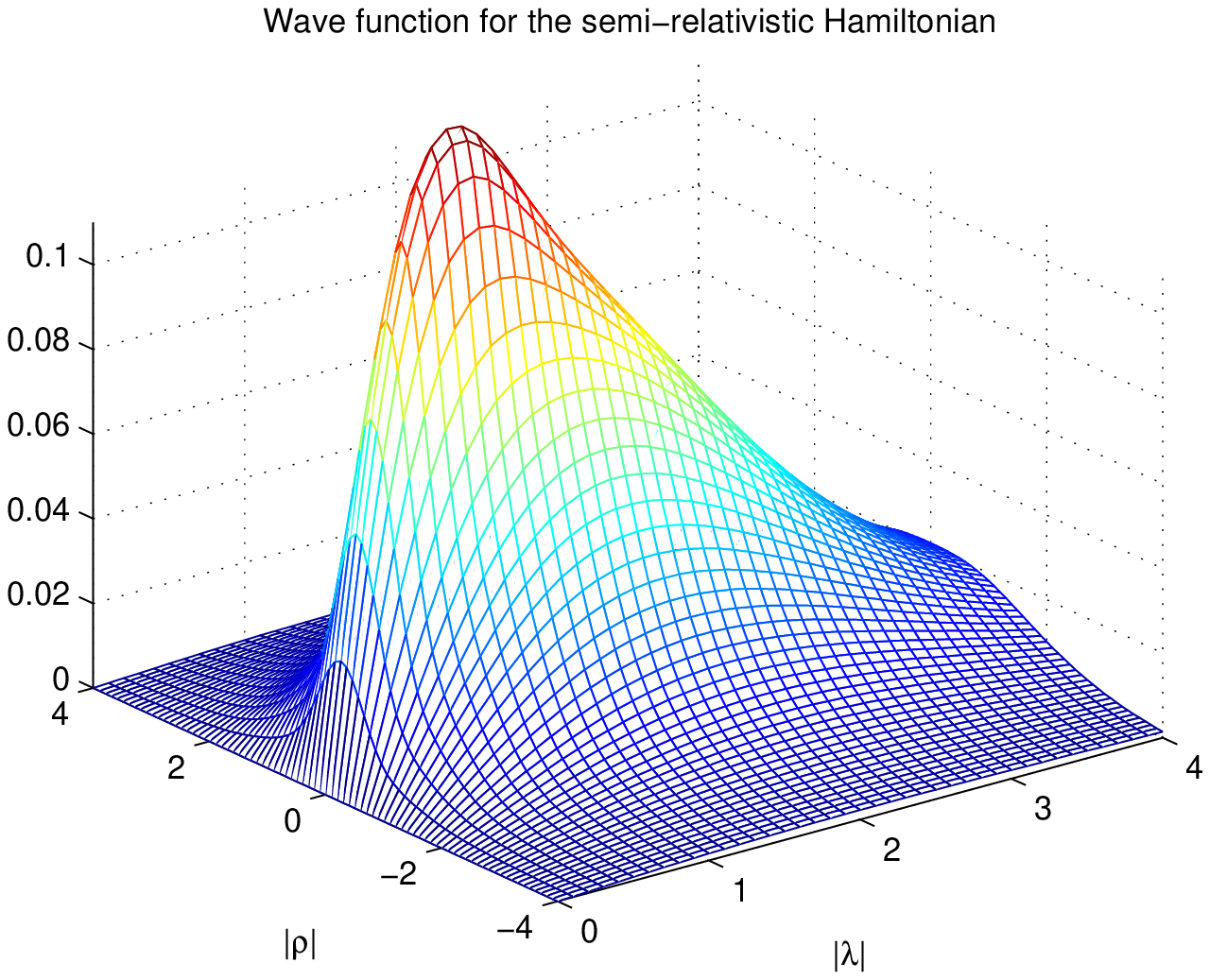}
  }
\caption{\label{fig:B:SR}Wave function for the semi-relativistic Hamiltonian
\eqref{eq:def:Gaussian} with 10 Gaussian functions \eqref{eq:gauss_gluon} for an electric
gluon hybrid meson.}
\end{center}
\end{figure}

\begin{table} [h]
\begin{ruledtabular}\caption{Spectra of \eqref{AF3} and \eqref{eq:mainH} with different wave functions. Parameters ~\eqref{eq:param} are used. All masses are in GeV.}
  \begin{tabular}{lccc}
     & Magn. ($A$) & Elec. ($B$) & Full\\
    \hline
    $H(\mu,\mu_g)$, $\chi(\bm X)$ \eqref{eq:trialfct_hyperS} & 4.573& 4.225 &\\
    $H(\mu,\mu_g)$, $N=1$  & 4.462& 4.137 & 4.137\\
    $H(\mu,\mu_g)$, $N\to\infty$  & 4.445& 4.115 & 4.068\\
    $H$, $N=1$  & 4.402& 4.017 & 4.017\\
    $H$, $N\to\infty$  & 4.361& 3.970 & 3.895\\
  \end{tabular}
  \label{tab:resul_H0}
\end{ruledtabular}
\end{table}

The final step is, of course, to diagonalize $H$ without any assumption on the wave function,
{\it i.e.} with \eqref{eq:ansatzfull}. The ground state mass decreases once again and we
obtain
\begin{equation}\label{}
M_0 = 3.895 \text{ GeV}.
\end{equation}
We summarize all results in Table~\ref{tab:resul_H0}, namely the spectra of $H(\mu,\mu_g)$ and
$H$ for magnetic ($A$) and electric ($B$) gluons, but also for a more general $P$-wave $c\bar
cg$ hybrid mesons (Full). In each case, we displayed the results for $N=1$ trial Gaussian
functions and the true values obtained typically with $N=10$ trial functions. In the general
$P$-wave cases, the $N=1$ give always the electric gluon mass since the latter is the main
component of the wave function. In view of this Table, we notice a gain of 600 MeV when
relaxing all hypothesis !

\begin{table} [htb]
\begin{ruledtabular}\caption{Wave function coefficients for magnetic gluon with the semi-relativistic Hamiltonian.}
  \begin{tabular}{cccc}
    $i$ & $\alpha_i$ $10^{-2}$& $a_i$ & $b_i$ \\
    \hline
    1 &3.138 &0.086588 & 0.205722 \\
    2 &6.262 &0.216618 & 0.781620 \\
    3 &0.179 &0.055116 & 0.070758 \\
    4 &$-0.287$ &0.608667 & 0.201197 \\
    5 &108.510 &0.875692 & 0.739526 \\
    6 &3.673 &0.161373 & 0.426376 \\
    7 &$-0.742$ &0.059422 & 1.098722 \\
    8 &$-2.084$ &0.339798 & 0.350189 \\
    9 &39.117 &1.054584 & 0.569019 \\
    10 &$-145.020$ &0.939776 & 0.679936 \\
  \end{tabular}
  \label{tab:wfmgtsemirelat}
\end{ruledtabular}
\end{table}

\begin{table} [htb]
\begin{ruledtabular}\caption{Wave function coefficients for electric gluon with the semi-relativistic Hamiltonian.}
  \begin{tabular}{cccc}
    $i$ & $\alpha_i$ $10^{-2}$& $a_i$ & $b_i$ \\
    \hline
    1 &3.672 &0.125160 &0.214905  \\
    2 &0.130 &0.059228 &0.058032  \\
    3 &89.281 &0.342251 &0.796071  \\
    4 &3.431 &0.580957 &0.262103  \\
    5 &$-0.666$ &0.150242 &1.634497  \\
    6 &$-8.155$ &0.491529 &0.358052  \\
    7 &21.727 &1.401183 &3.854743  \\
    8 &$-75.356$ &0.341822 &0.825713  \\
    9 &$-0.076$ &0.085946 &0.747611  \\
    10 &$ -3.160$ &2.586533 &0.989884  \\
  \end{tabular}
  \label{tab:wfelecsemirelat}
\end{ruledtabular}
\end{table}

\begin{figure}[htb]\begin{center}
\resizebox{0.5\textwidth}{!}{
  \includegraphics*{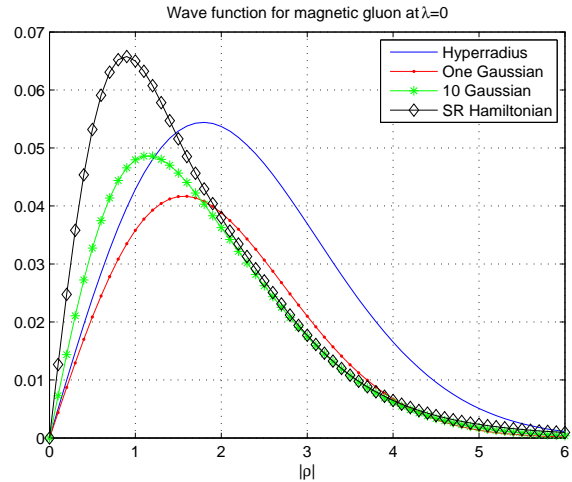}
  }
\caption{\label{fig:lambda0}Wave function for a magnetic gluon at $\lambda=0$.}
\end{center}
\end{figure}

\begin{figure}[htb]\begin{center}
\resizebox{0.5\textwidth}{!}{
  \includegraphics*{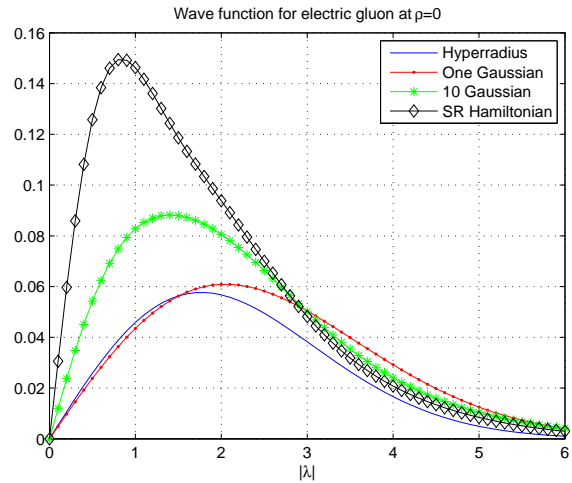}
  }
\caption{\label{fig:rho0}Wave function for an electric gluon at $\rho=0$.}
\end{center}
\end{figure}

In order to compare qualitatively the different wave functions, we display in Fig.~\ref{fig:lambda0} the wave functions for a magnetic gluon at $\lambda=0$ and in Fig.~\ref{fig:rho0} the wave functions for an electric gluon at $\rho=0$.

\begin{figure}[htb]\begin{center}
\resizebox{0.5\textwidth}{!}{
  \includegraphics*{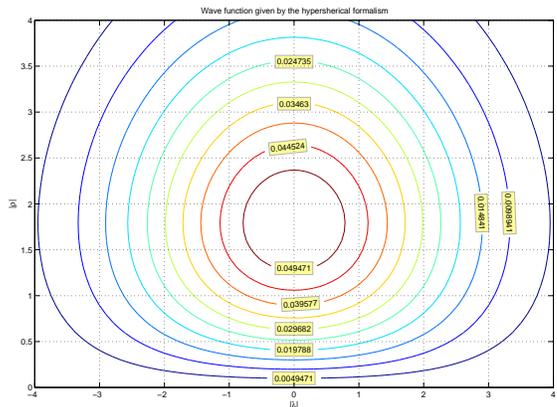}
  }\caption{\label{fig:A:Hyperplan}Wave function \eqref{eq:trialfct_hyperS} for a magnetic gluon hybrid meson.}
\end{center}
\end{figure}

\begin{figure}[htb]\begin{center}
\resizebox{0.5\textwidth}{!}{
  \includegraphics*{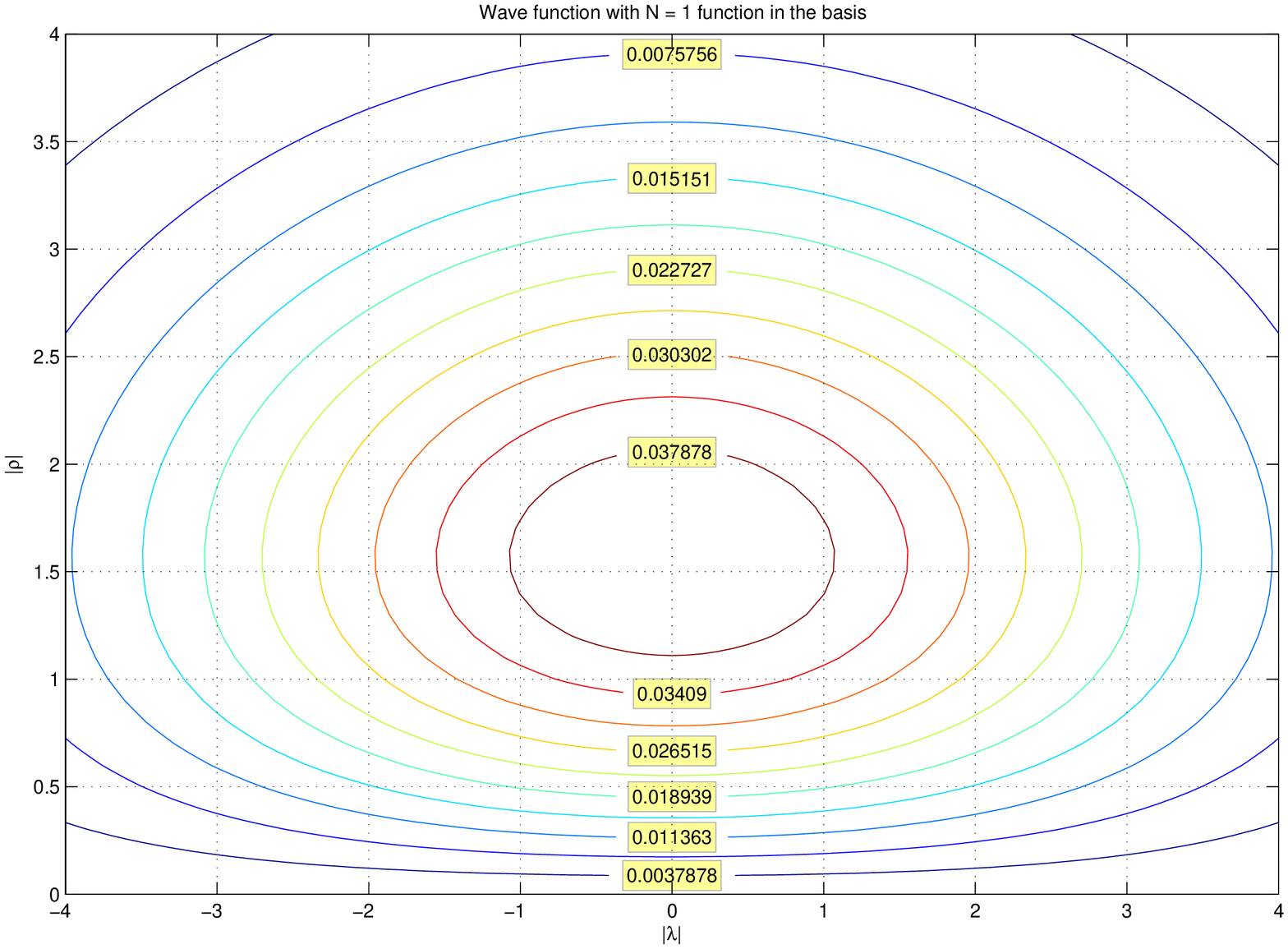}
  }\caption{\label{fig:A:oneplan}Wave function \eqref{eq:gauss_gluon} for a magnetic gluon hybrid meson.}
\end{center}
\end{figure}

\begin{figure}[htb]\begin{center}
\resizebox{0.5\textwidth}{!}{
  \includegraphics*{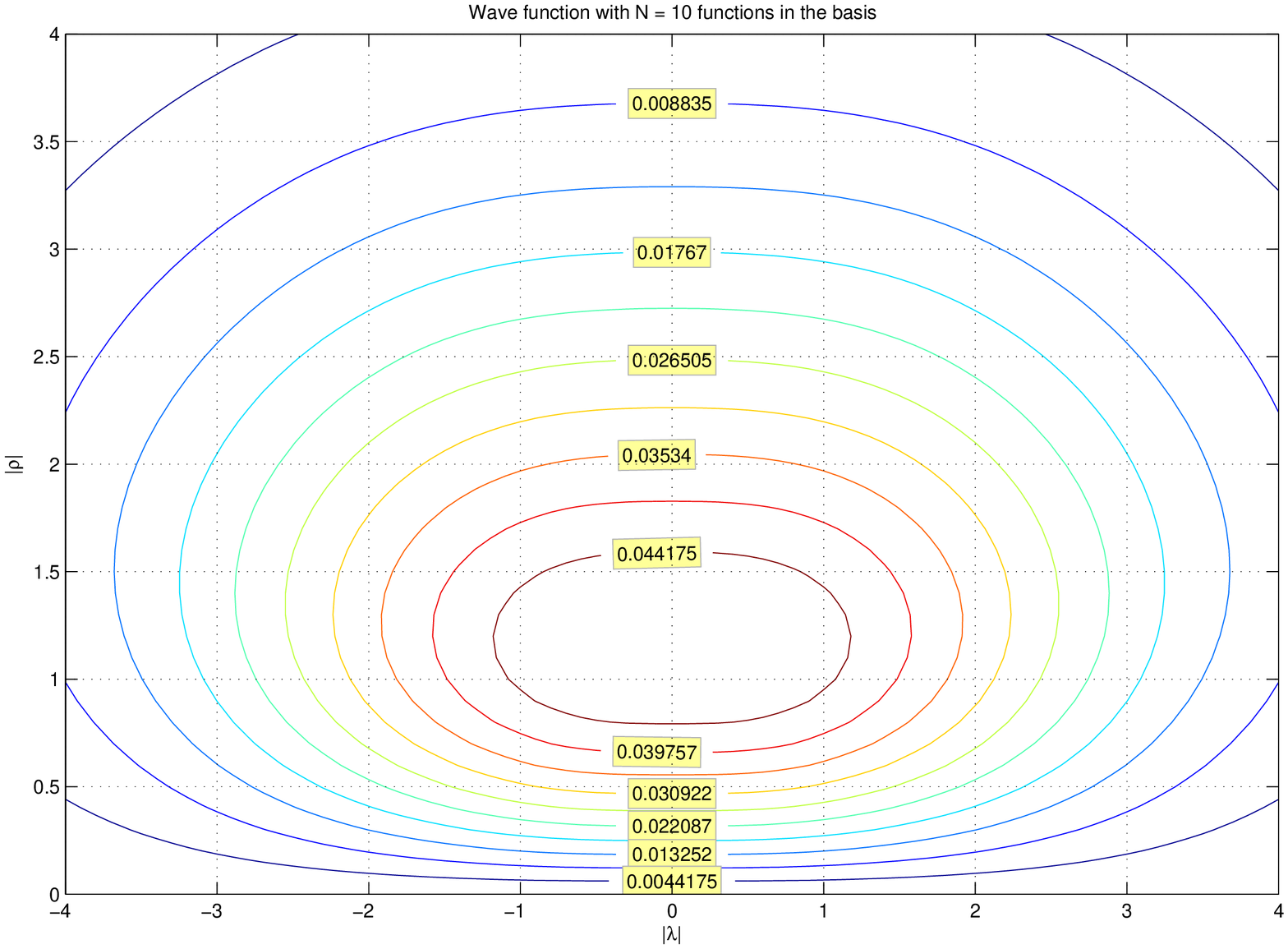}
  }
\caption{\label{fig:A:tenplan}Wave function \eqref{eq:def:Gaussian} with 10 Gaussian functions
\eqref{eq:gauss_gluon} for a magnetic gluon hybrid meson.}
\end{center}
\end{figure}

\begin{figure}[htb]\begin{center}
\resizebox{0.5\textwidth}{!}{
  \includegraphics*{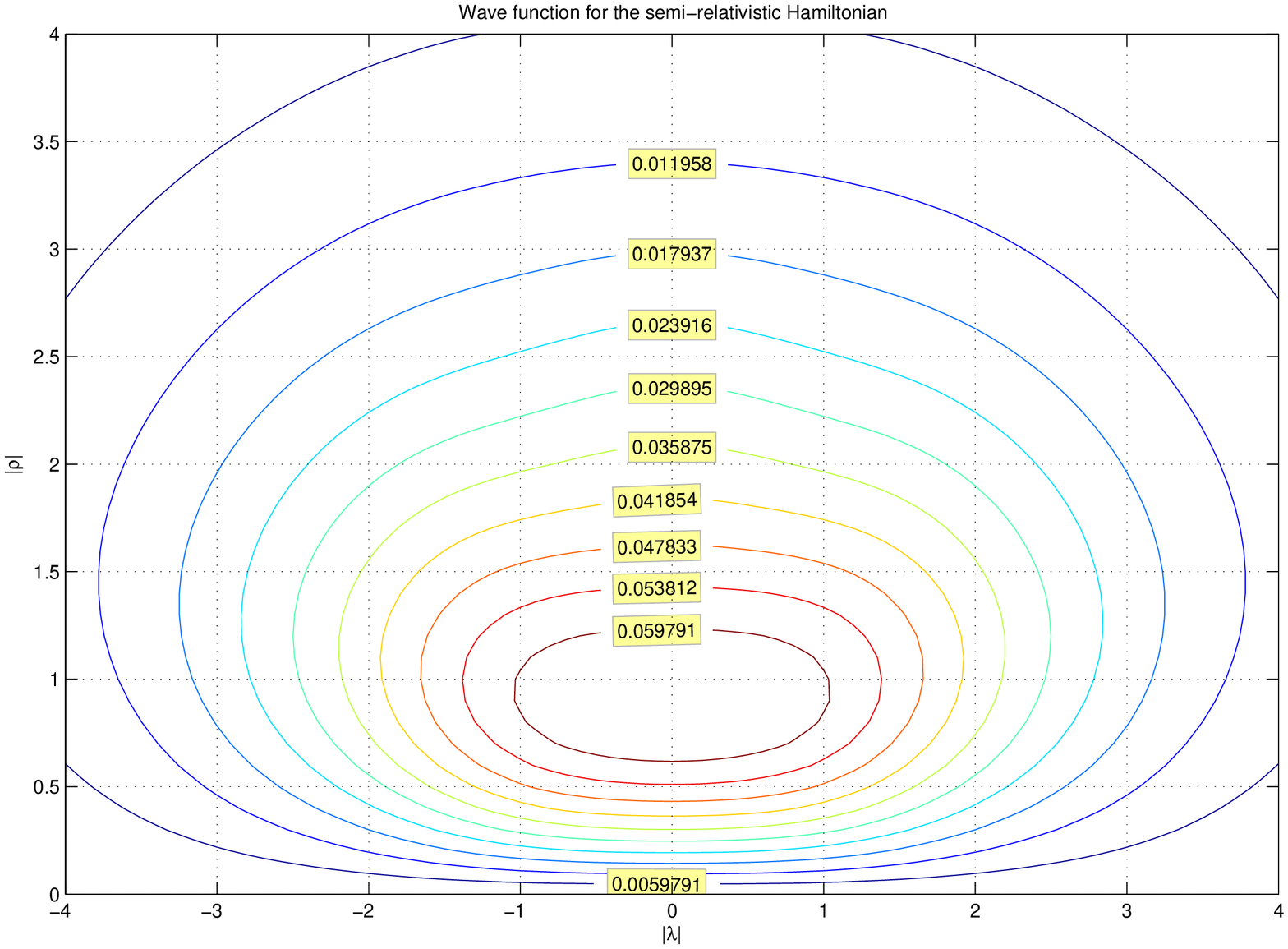}
  }
\caption{\label{fig:A:SR_plan}Wave function for the semi-relativistic Hamiltonian
\eqref{eq:def:Gaussian} with 10 Gaussian functions \eqref{eq:gauss_gluon} for a magnetic gluon hybrid meson.}
\end{center}
\end{figure}

\begin{figure}[htb]\begin{center}
\resizebox{0.5\textwidth}{!}{
  \includegraphics*{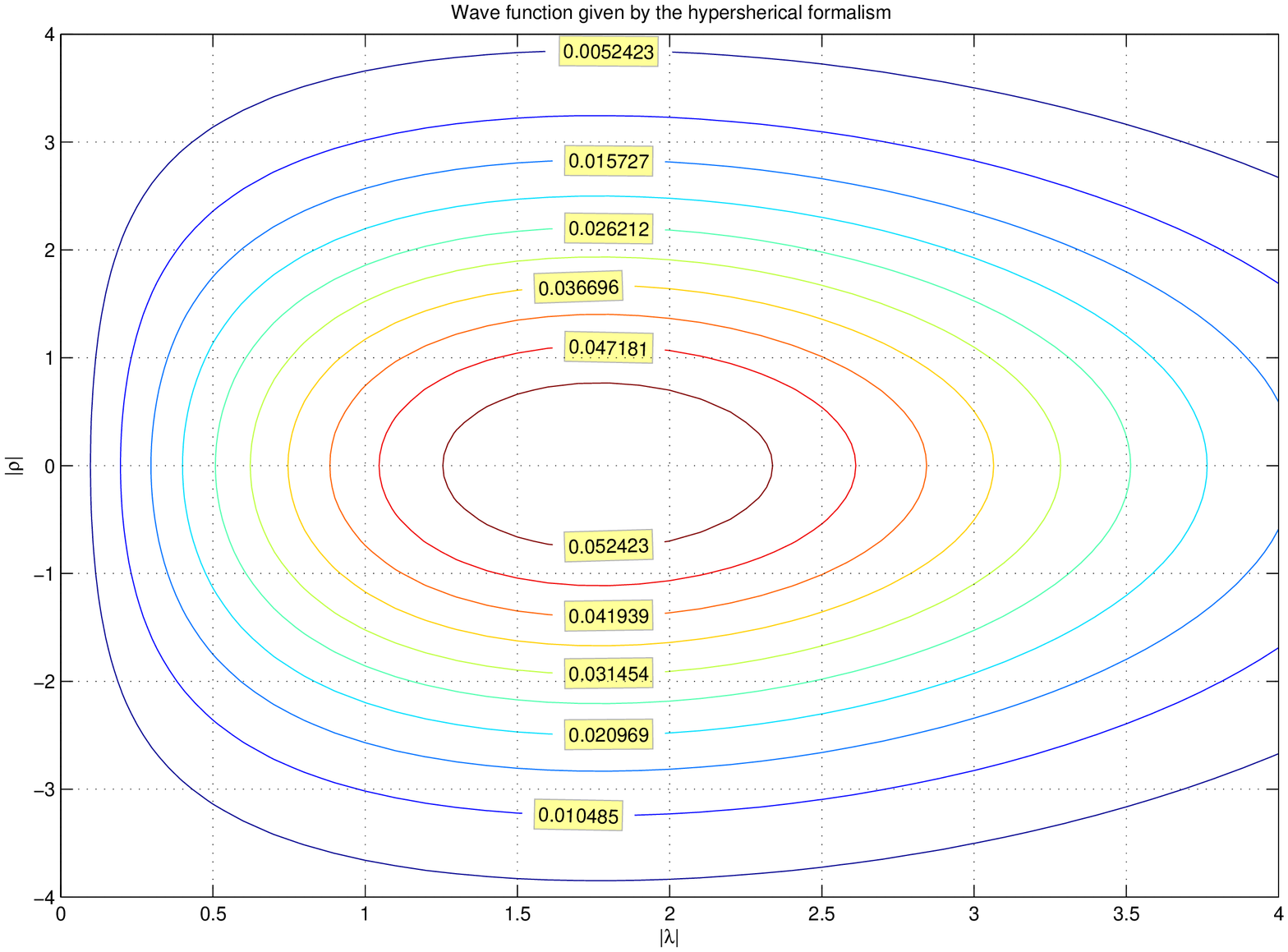}
  }\caption{\label{fig:B:Hyperplan}Wave function \eqref{eq:trialfct_hyperS} for an electric gluon hybrid meson.}
\end{center}
\end{figure}

\begin{figure}[htb]\begin{center}
\resizebox{0.5\textwidth}{!}{
  \includegraphics*{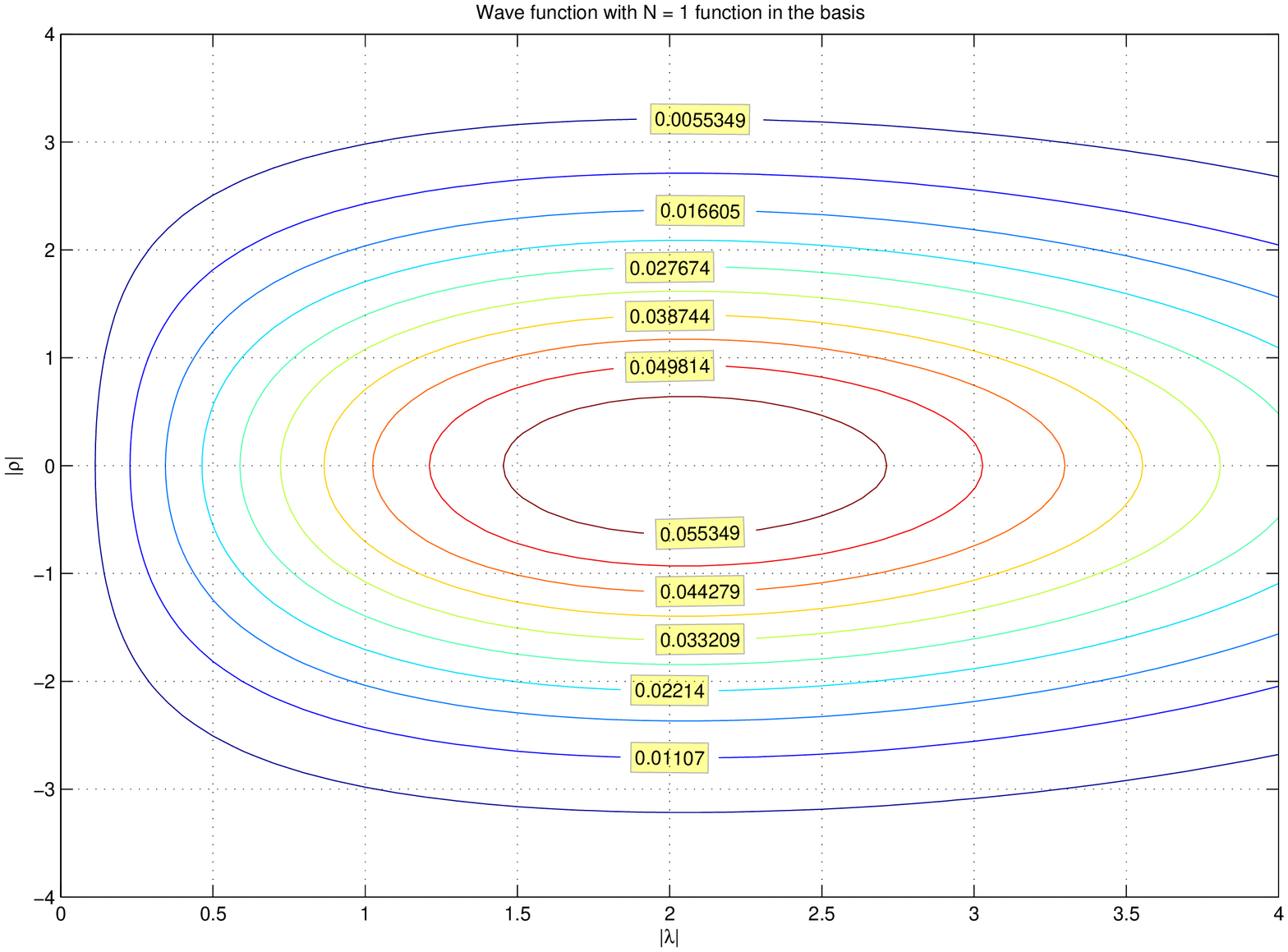}
  }\caption{\label{fig:B:oneplan}Wave function \eqref{eq:gauss_gluon} for an electric gluon hybrid meson.}
\end{center}
\end{figure}

\begin{figure}[htb]\begin{center}
\resizebox{0.5\textwidth}{!}{
  \includegraphics*{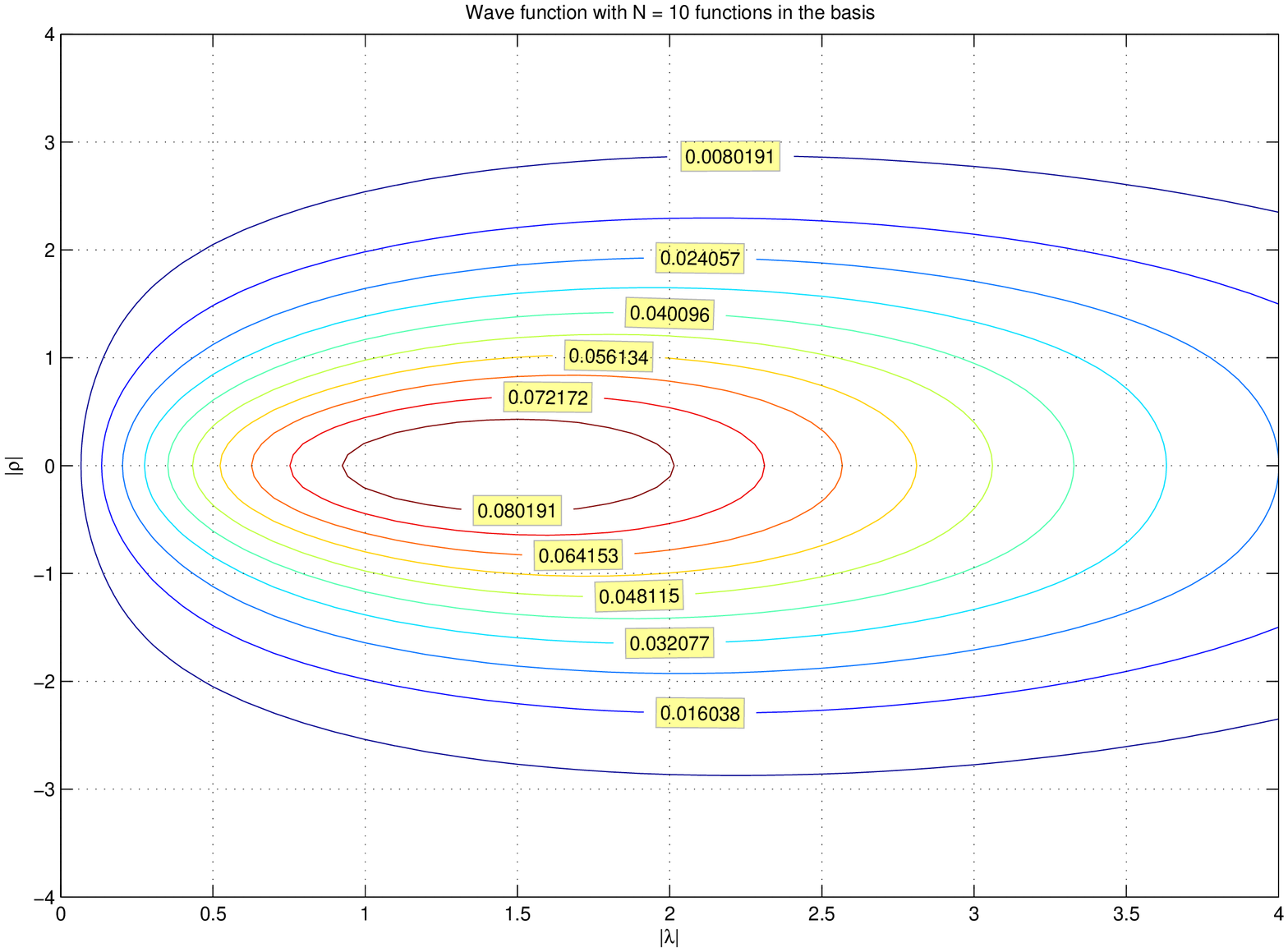}
  }
\caption{\label{fig:B:tenplan}Wave function \eqref{eq:def:Gaussian} with 10 Gaussian functions
\eqref{eq:gauss_gluon} for an electric gluon hybrid meson.}
\end{center}
\end{figure}

\begin{figure}[htb]\begin{center}
\resizebox{0.5\textwidth}{!}{
  \includegraphics*{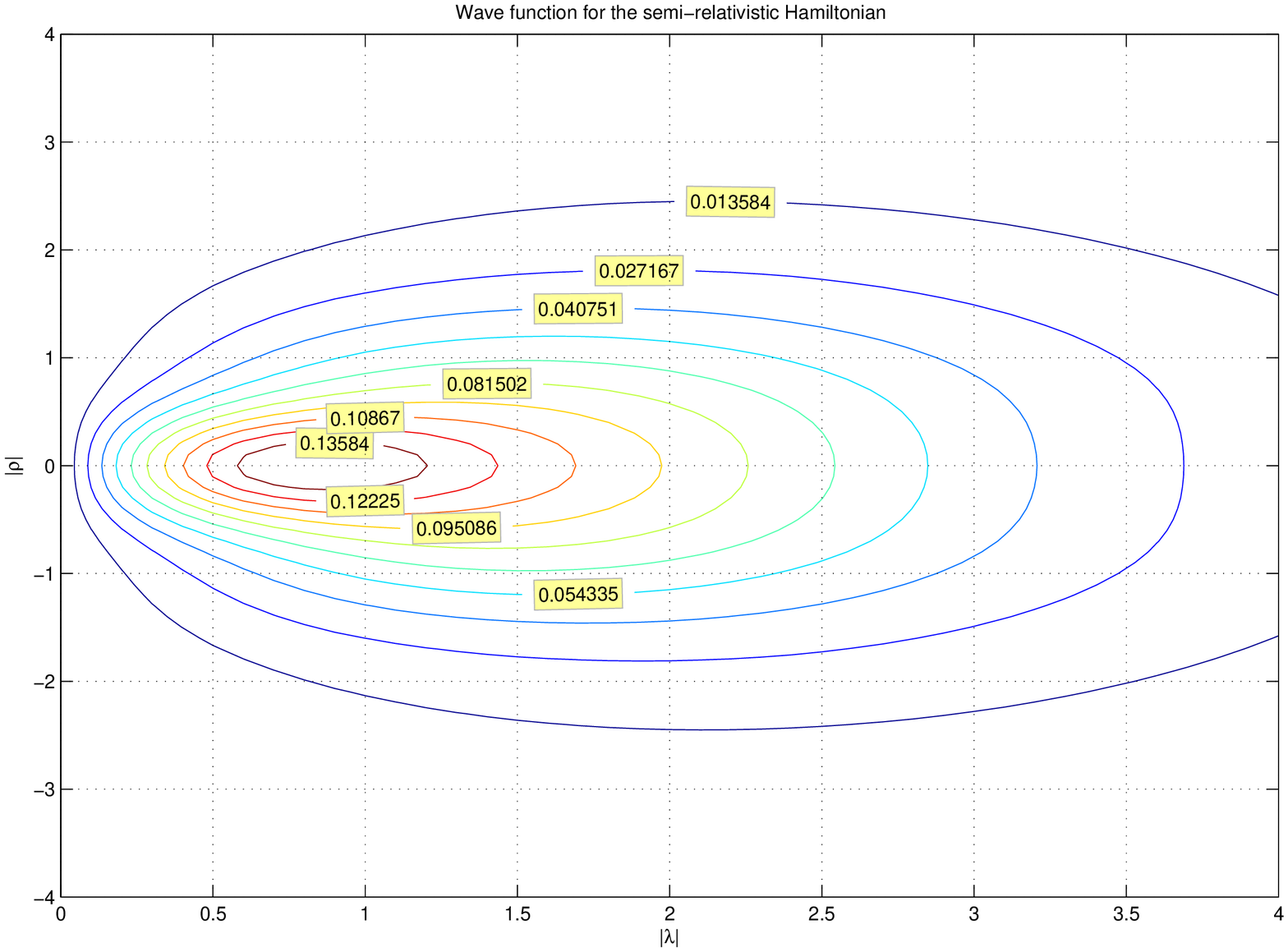}
  }
\caption{\label{fig:B:SR_plan}Wave function for the semi-relativistic Hamiltonian
\eqref{eq:def:Gaussian} with 10 Gaussian functions \eqref{eq:gauss_gluon} for an electric
gluon hybrid meson.}
\end{center}
\end{figure}

For the sake of completeness and in order to compare all the wave functions quantitatively, we add the contour lines of all the eight wave functions described in the text. Those plots are displayed in Figs.~\ref{fig:A:Hyperplan}, \ref{fig:A:oneplan}, \ref{fig:A:tenplan}, \ref{fig:A:SR_plan}, \ref{fig:B:Hyperplan}, \ref{fig:B:oneplan}, \ref{fig:B:tenplan} and \ref{fig:B:SR_plan}.

The mass of the hybrid with an electric gluon is resolved to be lighter than the magnetic one. Our simple Hamiltonian mixes the two states and the mass of the lightest states has clearly a bigger electric component. However, the decay modes of these two states are different. An electric gluon allows the hybrid to decay into a S-wave $D^*\bar{D}^*$. This may induce difficulty to single out a hybrid interpretation with respect to conventional charmonium.
I would be interesting to investigate more deeply the decay properties of those states which could guide experimentalist in the search for hybrid charmonia.

\section{Parameter Influence}\label{sec:param}
\begin{figure}[htb]\begin{center}
\resizebox{0.5\textwidth}{!}{
  \includegraphics*{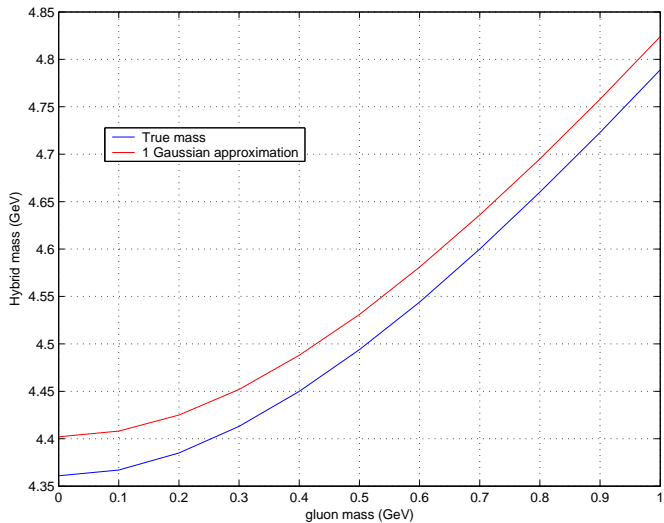}
  }
\caption{\label{fig:mass_evo}Hybrid mass evolution according to gluon mass.}
\end{center}
\end{figure}

\begin{figure}[htb]\begin{center}
\resizebox{0.5\textwidth}{!}{
  \includegraphics*{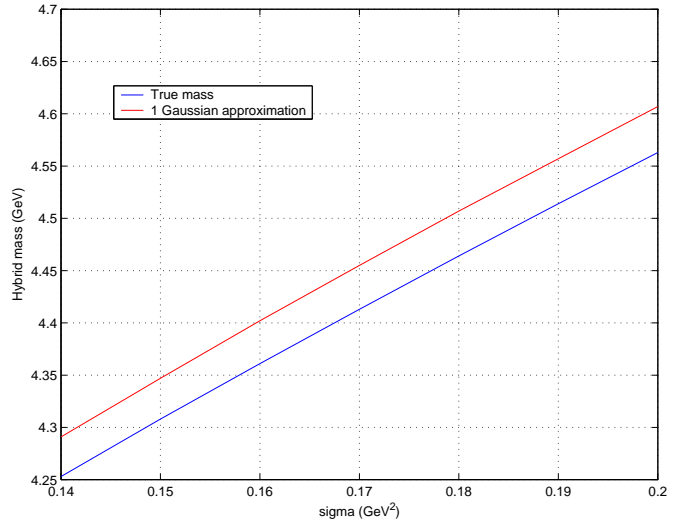}
  }
\caption{\label{fig:sigma_evo}Hybrid mass evolution according to string tension.}
\end{center}
\end{figure}

The previous sections emphasized on the fact that various approximations may lead to
overestimation of hybrid masses. However, the single Gaussian approximation was resolved to be an acceptable approximation. We now turn our attention to this specific approximation and test its evolution for different values of the parameters for the semi-relativistic
Hamiltonian~\eqref{eq:mainH}. We then change one parameter and keep the others to their
``optimal'' values \eqref{eq:param}. The results are displayed in Figs~\ref{fig:mass_evo} and
\ref{fig:sigma_evo} for respectively the evolution of the gluon mass and the string tension.
Varying the gluon mass is interesting since effective approaches used a non-vanishing value
for the gluon mass in the kinetic energy, see for instance Ref.~\cite{Szczepaniak:2005xi}. For wide ranges of the parameters, the single Gaussian approximation is resolved always to give an overestimation of $\sim 40$ MeV constant in the intervals. For a wide range of variation of
$\alpha$, the single Gaussian approximation lie around 20-50 MeV above the true mass. For
hybrid meson systems, the approximative wave function \eqref{eq:gauss_gluon} is then robust
under parameter evolution. We checked that we obtain similar results for electric gluon wave
functions.

\section{Conclusion}\label{sec:conclu}

During the main part of this paper, we were concerned by Kalashnikova and Nefediev's model
\cite{Kalashnikova:2008qr} in which the starting point was Hamiltonian \eqref{eq:mainH}.
This model shares the constituent model's typical features. The kinetic energy is the
semi-relativistic expression valid for massless particles (the constituent gluon). A linear
(in fact a V-junction) plus Coulomb term is used for the potential. The parameters (charm
quark mass, string tension and strong coupling) were determined on a similar model for
charmonium. The parameters \eqref{eq:param} reproduce a quarkonium spectrum in agreement with
experimental data and are expected to predict an acceptable value for hybrid masses.

However, solving a three-body systems may require hypothesis: On one side in the Hamiltonian
by introducing auxiliary fields to get of rid of square roots; on the other side on the wave
function by imposing particular values for internal quantum numbers and/or restricting the
space spanned by the wave function to a restricted subspace of the total Hilbert space.

We demonstrated in this work that one should be very careful with the approximations made to
solve the eigenvalue problems numerically. Indeed, auxiliary fields, hyperspherical formalism
with only one function, Fock space reduction (magnetic/electric gluons) may causes an
overestimation of the mass by amount of 100-600 MeV. However, in each cases of study (magnetic
or electric gluon), the less worst approximation was the only Gaussian function. For the same
Hamiltonian, the single Gaussian overestimated the mass by only $\sim 30$ MeV where the
hyperspherical formalism with only one function gave always a discrepancy around $130$ MeV.

In their conclusion, the authors of \cite{Kalashnikova:2008qr} pointed out a numerical vector
hybrid meson mass of 4.397 GeV, substantially higher than the experimental candidate
$X(4260)$. The improvements of the numerical method developed in the present paper clearly
favor the hybrid meson interpretation of the candidate. Indeed, the real eigenvalue of the
Hamiltonian \eqref{AF3} used in \cite{Kalashnikova:2008qr} is 130 MeV below the approximation
with the hyperspherical formalism, the exact discrepancy with the vector X(4260). We stressed, nevertheless, that the correction to the bare mass were computed with the approximative wave
function and should slightly differ. It would be interesting to investigate how strong is the
difference. Moreover, we have shown that the mass remains more or less stable under the
approximations considered in this paper but we think that the decays properties, depending
strongly of the wave functions, should change more that the mass.

It is also worth mentioning the two other vector states Y(4325) from BaBar~\cite{Aubert:2006ge} and Y(4360) from Belle~\cite{wang:2007ea}. Their mass lie close to the X(4260) and mixing may cause a mass shift with respect to pure $c\bar cg$ hybrid states. However, the purpose of this paper was the investigation of the numerical procedure for a simple Hamiltonian for pure $c\bar cg$. A detail investigation of the possible mixing with other states would require a more complicated description of the system and is beyond this study.

\section*{Acknowledgements}
The author thanks F. Buisseret and C. Semay for valuable suggestions and comments about this
work and the I.I.S.N. for financial support.




\end{document}